# Study of ion beam induced mixing in nano-layered Si/C multilayer structures

Ram Prakash[*], S. Amirthapandian[$], D. M. Phase[*@], S. K. Deshpande[#],

R. Kesavamoorthy[$] and K.G. M. Nair[$]

[*]UGC - DAE Consortium for Scientific research, Indore Centre, University campus, Khandwa road, Indore – 452 017, India.
[$]Materials Science Division, Indira Gandhi Centre for Atomic Research,
Kalpakkam-603 013, India
[#]UGC-DAE Consortium for Scientific research, Mumbai Centre, R5 Shed, Bhabha Atomic Research Centre, Trombay, Mumbai-400 085, India.

*Abstract*

The effects of ion beam induced atomic mixing and subsequent thermal treatment in Si/C multilayer structures are investigated by use of the technique of grazing incidence X-ray diffraction (GIXRD) and Raman spectroscopy. The [Si (3.0 nm) / C (2.5 nm)]$_{\times 10}$/Si multilayer films were prepared by electron beam evaporation under ultra high vacuum (UHV) environment. The layer thicknesses were measured using *in-situ* quartz crystal oscillator. These multilayer films were subjected to 40 keV $Ar^+$ ion irradiation with fluences $5 \times 10^{16}$ (low fluence) and $1 \times 10^{17}$ ions/$cm^2$ (high fluence). The as-prepared and irradiated multilayer samples were annealed at 773 K for one hour. The GIXRD and Raman spectroscopy results reveal the formation of different phases of SiC in these multilayer structures. Deposition induced reactions at the nano-structured interface and subsequent room temperature Ar ion irradiation at low fluence result in formation of the hexagonal SiC phase. High fluence $Ar^+$ ion irradiation and subsequent annealing at 773 K for one hour leads to precipitation of the cubic-SiC phase.

[@]Corresponding author e mail : dmphase@udc.ernet.in

# 1. Introduction

Silicon carbide (SiC) is an important non-oxide ceramic material with a combination of unique physical, chemical, electronic and mechanical properties [1]. As a wide band semiconductor, it is an excellent candidate for high temperature, high power electronic devices. Due to its high resistance to oxidation, corrosion, wear and creep it has enormous applications in tribology [2].

Conventional procedures for producing crystalline SiC require very high temperature. A solid-state reaction between carbon and silicon does not occur below 2273 K. So, to develop a SiC coating technique on silicon at moderate temperature, which is compatible with existing Si processing techniques, is a topic of current interest. Several techniques have been used to form SiC coatings such as chemical vapour deposition (CVD) [3], ion beam assisted deposition (IBAD) [4], molecular beam epitaxy (MBE) [5], ion implantation and ion beam mixing (IBM) [6,7]. IBM is a local rearrangement of atoms across the interface induced by energetic ions [8]. It is a processing tool to synthesize metastable phases such as solid solutions from miscible and/or immiscible elements [9]. A major advantage of IBM over ion implantation technique is its ability to produce materials with higher solute concentration at lower irradiation fluence. There are few attempts to synthesise SiC from Si/C multilayers using low energy ion beams [7,10] by other groups. However, in those studies individual layer (Si and C) thicknesses in the multilayer structure were thick (>5nm) and it was found that complete mixing and crystalline phase formation was achieved by irradiation at high temperature (873 K) only. Complete mixing at low temperatures was also achieved, but the obtained SiC phase was amorphous.

Very recently some IBM studies [11] have demonstrated that when the layer thickness was reduced to a couple of nanometers, the interfacial free energy could be quite high and could drive the multilayers to some previously unknown energetic states, helping to explore new metastable states. In the present paper, we report upon ion beam mixing in Si/C multilayer structures having individual layer thicknesses (of Si and C) in 2-3 nanometer range.

The compound formation and complex nature of defects created by ion bombardment can be studied by grazing angle X-ray diffraction. Raman Scattering is also a powerful tool to probe the implanted region because the penetration depth of laser is of the order of implantation-affected surface thickness [12]. Raman spectroscopy is used to get information on large variety of crystalline parameters of semiconducting materials [13]. Combined grazing incidence x-ray diffraction and Raman scattering measurements are used to characterise as-prepared, ion beam mixed and subsequently annealed samples and different phases of SiC are identified.

## 2. Experimental Details

The [Si (3.0 nm) / C (2.5 nm)]$_{\times 10}$/Si multilayer films were prepared by using an ultra high vacuum electron beam evaporation system developed at our laboratory [14]. The deposition was carried out at a background pressure of $2 \times 10^{-7}$ Pa. The film thickness was monitored using a quartz crystal monitor. The deposition rate was maintained at 0.01 nm/s. The room temperature 40 keV Ar ion irradiation was carried out on Si/C multilayer films at fluences of $5 \times 10^{16}$ ions/cm$^2$ (low fluence) and $1 \times 10^{17}$ ions/cm$^2$ (high fluence) using a 150 kV ion implanter at IGCAR, Kalpakkam. The vacuum inside the chamber was of the order of $6 \times 10^{-5}$ Pa during ion irradiation. In order to avoid beam heating, the beam current density was limited to less than 0.5 µA/cm$^2$. The calculated value of the projected range plus straggling ($R_p + \delta R_p$) of 40 keV Ar$^+$ ions using the SRIM code [15] is about 58 nm, which is roughly equal to thickness of the multilayer structure. The Ar$^+$ ion energy is chosen to maximise the damage production in the film. The as-prepared and irradiated

Si/C multilayer films were annealed at 773 K for one hour at the vacuum of 6 x $10^{-5}$ Pa. GIXRD pattern of as-prepared, ion irradiated and subsequent annealed samples were recorded using Siefert, machine (Model: XRD -3003TT). Cu-$K_\alpha$ radiation was used and the scan was done in a 2θ range of 20° to 75°. The angle of incidence was kept at 0.2° in order to eliminate substrate contributions. Raman spectra were recorded at room temperature using a 514.5 nm argon laser beam of 600 mW power focussed to a spot size of 50 μm. The scattered light was collected in backscattering geometry using a camera lens (Nikkon, focal length 5cm, f/1.2) and a focussing lens. The collected light was dispersed in a double grating monochromator (SPEX model 14018) and detected using a thermoelectrically cooled photomultiplier tube (model ITT-FW 130). Raman spectra were recorded from 500 to 540 $cm^{-1}$ (Si modes) and from 700 to 1100 $cm^{-1}$ (SiC modes) at interval of 1 $cm^{-1}$ and collection time of 10 s at each step. The instrument resolution was 4.5 $cm^{-1}$. These spectra were fitted using different Gaussian peaks in order to separate the contributions of different phases.

### 3. Results and Discussion

Figure 1 shows the GIXRD pattern corresponding to as deposited, ion beam mixed and subsequent annealed Si/C multilayer samples. The observed peak position and corresponding phases are tabulated in Table-1. The spectrum corresponding to the as deposited sample (Fig.1(a)) shows major contribution of silicon and a small contribution of the hexagonal phase of SiC. Since in our multilayer structures the individual layer thicknesses are small, formation of metastable SiC phase is attributed to deposition induced reactions. The absence of the SiC peaks in the spectrum corresponding to the annealed sample (Fig.1(b)) indeed confirm the metastability of this phase. The spectrum corresponding to the low fluence irradiated sample (Fig.1(c)) shows the same feature as observed in that of the as-prepared sample. However, upon thermal treatment the low fluence irradiated sample shows a different behaviour when compared with as-prepared and annealed sample. The spectrum (Fig.1(d)) shows an increase of the contribution of the hexagonal phase of SiC. This difference is attributed to the presence of defects in the irradiated sample, which enhances the inter diffusion and solid state reaction across the interface upon thermal treatment. Further bombardment at high fluence results in a drastic change of the GIXRD spectrum (Fig.1(e)). When compared with spectra corresponding to as deposited (Fig.1(a)) and low fluence irradiated sample (Fig.1(c)), this spectrum shows the absence of peaks corresponding to silicon. This confirms that complete mixing has occurred in the multilayer structure. The spectrum shows emergence of new peaks corresponding to hexagonal and cubic SiC phases. The additional diffraction peak at $22.4^o$ matches well with carbon cluster. Also, all the diffraction peaks corresponding to SiC phase show small variation in $2\theta$ position when compared with standard powder diffraction data. This is expected in our case as a high density of defects are produce in the sample due to high fluence Ar ion bombardment. Besides the defect generation these argon atoms are physically present in the mixed region. Though they are chemically inert but their atomic diameter is larger than that of Si and C which leads to the stresses in ion bombarded sample. These stresses are mainly responsible for variation in $2\theta$ position in GIXRD spectrum. Upon annealing at 773 K the spectrum corresponding to the high fluence irradiated sample shows the presence of peaks corresponding to cubic SiC and silicon. This shows that upon thermal treatment at moderate temperature, silicon atoms tend to outmigrate from the defective state produced due to high fluence ion bombardment. This outmigration of silicon may be attributed to the presence of higher amount of silicon as compared to carbon (as measured by quartz crystal monitor during deposition) in our deposited multilayer structure. In order to maintain the stoichiometry of SiC phase excess silicon may have come out.

Figure 2 shows the Raman spectra recorded for the as-prepared, irradiated and subsequent annealed Si/C multilayer sample in the range of 500 to 540 cm$^{-1}$. The as-prepared multilayer sample (Fig.2(a)) shows Raman modes at 509, 520, and 522 cm$^{-1}$. The Raman mode at 509 cm$^{-1}$ corresponds to 6H-SiC [16], 520 cm$^{-1}$ is from Si substrate and 522 cm$^{-1}$ is from unmixed silicon layer [17,18] in the multilayer. The reduction in the Raman shift for the silicon layer is due to defects present in the silicon layer, which is expected due to strain in thin films. The annealed multilayer shows (Fig.2(b)) Raman modes at 518 and 522 cm$^{-1}$. Annealing leads to the disappearance of 520 cm$^{-1}$ mode, mainly due to reduction in defects. For the irradiated multilayer with fluence of 5 x 10$^{16}$ ions / cm$^2$ (Fig.2(c)), Raman modes are observed at 508, 520, 524 and 537 cm$^{-1}$. The mode at 508 cm$^{-1}$ corresponds to 6H-SiC [16] and the other two modes to the silicon substrate (520 cm$^{-1}$) and silicon layer (524 cm$^{-1}$) present in multilayer. The mode at 537 cm$^{-1}$ is due to strain present in the silicon layer. After annealing at 773 K the modes at 508 and 537 cm$^{-1}$ disappeared (Fig.2(d)) and other shifted to 518 and 522 cm$^{-1}$, i.e. annealing suppresses the 6H-SiC phase formation and defects are annealed out. The increase of the area of the Raman peaks after annealing also confirm the reduction of stresses in the silicon layer as well as silicon substrate. The sample irradiated with the high fluence of 1×10$^{17}$ ions / cm$^2$ (Fig.2(e)) shows the Raman modes at 518 and 522 cm$^{-1}$ and after post annealing (Fig.2(f)) these modes are shifted to 520 and 524 cm$^{-1}$ Hence annealing plays an important role in the formation of the stiochiometric phase free from any defect which are generated during irradiation. It is clearly seen that irradiation leads to increase in strain where as annealing releases strains.

Figure 3 shows the Raman spectra recorded in the range from 700 to 1100 cm$^{-1}$. The as-prepared multiplayer (Fig.3(a)) shows Raman modes at 735, 763, 836, 930, 968, 981, 1051 and 1075 cm$^{-1}$. The Raman mode at 763 cm$^{-1}$ corresponds to Folded Transverse

Optical (FTO) phonon mode of 6H-SiC [19]. The other modes indicate the presence of 3C-SiC, and 15R-SiC phases [17,20,21]. The sample annealed at 773 K (Fig.3(b)) shows Raman modes at 770, 800, 883, 966, 977, 1066 cm$^{-1}$. The Raman modes at 770 and 883 cm$^{-1}$ correspond to the FTO mode of the 6H-SiC phase [20] and the modes at 800 and 977 cm$^{-1}$ correspond to Transverse optical (TO) [17] and Longitudinal optical (LO) [21] modes of the 3C-SiC phase. The mode appearing at 1066 cm$^{-1}$ may be due to defects. Annealing leads to the formation of 15R-SiC, where which is suppressed by irradiation as well as post annealing on irradiated samples.

The sample irradiated with 5 x 10$^{16}$ ions/cm$^2$ (Fig.3(c)) shows Raman modes at 772, 866, 933, 969, 988 and 1067 cm$^{-1}$. The Raman mode at 772 cm$^{-1}$ corresponds to the FTO mode of 15R-SiC [20] and the modes at 866 and 933 cm$^{-1}$ correspond to Folded Longitudinal Optical (FLO) modes of 15R-SiC [20]. The Raman mode at 969 cm$^{-1}$ corresponds to the LO mode of 3C-SiC [16]. The modes at 988 and 1067 cm$^{-1}$ may be due to defects present in the sample produced by irradiation. After annealing (Fig.(3(d)) these modes were shifted to 831, 930, 969, 977, 1046 and 1084 cm$^{-1}$. The Raman mode at 831 cm$^{-1}$ represents the FLO mode of 6H-SiC [16] and the mode 930 cm$^{-1}$ represents the FLO mode of 15R-SiC [20] while the mode at 969 [17] and 977 cm$^{-1}$ [21] correspond to the LO mode of 3C-SiC. The Raman modes at 1046 and 1084 cm$^{-1}$ may be due to defects. Since irradiation induces defects but annealing reduces stresses and defects in the sample the area and height of each peak is increased after annealing.

The sample irradiated up to the fluence of 1 x 10$^{17}$ ions/cm$^2$ (Fig.3(e)) shows Raman modes at 849, 934, 971, 980, 1046 and 1072 cm$^{-1}$. The modes at 849 and 931 cm$^{-1}$ correspond to the 15R-SiC phase [20] and the modes at 971 and 980 cm$^{-1}$ correspond to the 3C-SiC phase [18,21]. After post annealing of the irradiated sample (Fig.3(f)) Raman modes were observed at 931, 975, 990, 1029 and 1078 cm$^{-1}$. The mode at 931 and 975 cm$^{-1}$ correspond to 15R-SiC [20] and 3C-SiC [18] respectively. Annealing increases the area of observed modes.

In the present work, we found 6H-SiC, 15R-SiC and small volume fraction of 3C-SiC in the as-prepared Si/C multilayer itself. Heat of formation for 6H-SiC and 15R-SiC may be much smaller than for 3C-SiC and comparable with energy deposition during film growth. It may be the reason for the small volume fraction of 3C-SiC as observed from the Raman spectrum and might be due to the small layer thicknesses of Si and C layer (leading to the large interfacial free energy) as well as to the deposition method i.e. electron beam evaporation in which energy deposition during growth used to be in the range of 0.2 to 0.4 eV/atom. The 40 keV $Ar^+$ ion irradiation leads to the growth of 3C-SiC as evident from the LO phonon mode of 3C-SiC observed with Raman spectroscopy. According to the fractal geometry approach of Cheng [22], spikes are formed in systems having mean atomic number ($\bar{z}$) greater than 20. However in a detailed investigation of ion beam induced transport through bi-layer interfaces of low and medium Z metals, W.Bolse [23] reported diffusion in local thermal spikes as a major mixing mechanism. The average atomic number of Si/C system is 10, which is less than 20, hence initiation of local thermal spikes upon Ar ion bombardment can not be ruled out. According to the empirical formula proposed by Rossi *et al*. [24] a linear dependence exists between the critical temperature $T_c$ for the onset of radiation enhanced diffusion and the average cohesive energy ($\Delta H_{coh}$). Using the empirical relation $T_c = 95.2\ \Delta H_{coh}$ (eV/atom), for the Si-C system, the $T_c$ was calculated to be 571 K. Hence mixing due to RED is ruled out as we have done the irradiation at room temperature. In our view the formation of 3C-SiC phase upon high fluence Ar ion irradiation is mainly a two step process : 1.Nucleation due to high interfacial free energy in deposited multilayer structure and 2. Short range diffusion leading to growth due to formation of local thermal spikes upon ion bombardment..

## 4. Conclusion

The Si/C multilayer samples were prepared and characterised by XRD and Raman Scattering before and after ion-irradiation and subsequent thermal annealing. Even in the as-prepared sample the XRD and Raman Scattering show the formation of a small volume fraction of SiC. It is shown that with the important contribution from the interfacial free energy, properly designed Si-C multilayer structures were able to undergo a short range diffusion upon ion irradiation at room temperature. The low fluence irradiation favours the formation of hexagonal phase while high fluence irradiated sample upon thermal treatment at 773 K result the precipitation of cubic SiC phase.

**Acknowledgements**

Authors from UGC-DAE CSR wish to thank Prof. V.N.Bhoraskar, Prof. Ajay Gupta and Dr.P.S.Goyal for encouragement. They also thank Mr. S. R. Potdar for help rendered in sample preparation. One of the authors (RP) is gratefully acknowledged the CSIR, New Delhi for fellowship.

**References**

1. H. Morkoc, S. Strite, G. B. Gao, M. E. Lin, B. Sverdlow and M. Burns, J. Appl. Phys. 76 (1994) 1363.
2. B. Szeptycka, A. Gajewska, Solid State Phenomena, 94 (2003) 245.
3. K. C. Kim, C.I. Park, J.I. Roth, K. S. Nahm, Y. H. Seo, J. Vac. Sci. Technol. A 19 (2001) 2636.
4. M. Zayotouni, J. P. Riviere, M. F. Denanot and J. Allain, Thin Solid Films 287 (1996) 1.
5. C. Guedj, M. W. Dashiell, L. Kulik, J. Kolodzey, J. Appl. Phys., 84 (1998) 4631.
6. F. Eichhorn, N. Schell, W. Matz, R. Kogler, J. Appl. Phys. 86 (1999) 4184.
7. F. Harbsmeier, W. Bolse, A. M. Flank, Nucl. Instr. and Meth. B, 166-167 (2000) 385.
8. M. Nastasi , J. M. Mayer, Mat. Sci. Eng. R 12 (1994) 1.
9. S. Amirthapandian, B. K. Panigrahi, A. K. Srivastava, Ajay Gupta, K. G. M. Nair, R. V. Nandedkar, and A. Narayanasamy J. Phys.: Condens. Matter 14 (2002) L641.
10. J. P. Riviere, M. Zayotouni, M. F. Denanot and J. Allain, Mat. Sci. Engg. B, 29 (1995) 105.
11. Z.C.Li, L.T.Kong and B.X.Liu, J.Phys.: Condens. Matter 14 (2002) 1.
12. C. S. R. Rao, S. Sundaram, R. L. Schmidt and J. Comas, J. Appl. Phys. 54 (1983) 1808.
13. J. Jimenez, E. Martin, and A. C. Prieto, Mater. Lett. 12 (1991) 132.
14. S.M. Chaudhari, N. Suresh, D.M. Phase, A. Gupta, B.A. Dasannacharya, J. Vac. Sci. Technol. A17 (1999) 242.
15. J. F. Ziegler, J.P.Biersack, U. Littmark, The Stopping and Ranges of Ions in Solids, Pergamon , New York, 1985.
16. T. Tomita, S. Saito, M. Baba, M. Hundhausen, T. Suemoto and S. Nakashima. Phys. Rev. B 62 (2000) 12896.
17. Z. C. Feng, C. C. Tin, R. Hu and J. Williams, Thin Solid Films 266 (1995) 1.
18. C. Serre, A. Perez-Rodriguez and J. R. Morante, J. Appl. Phys. 77 (1995) 2978.
19. T. Rajagopalan, X. Wang, B. Lahlouh, C. Ramkumar, P. Dutta and S. Gangopadhyay, J. Appl. Phys. 94 (2003) 5252.
20. S. Nakashima, H. Harima, T.Tomito and T. Suemoto, Phys. Rev.B 62 (2000) 16605.
21. D. N. Talwar and Z. C. Feng, Comput. Mater. Sci.30 (2004) 419.
22. Y. T. Cheng, Mater. Sci. Rep. 5 (1990) 45.

23. W.Bolse, Mat.Sci.Eng.Rep.12 (1994) 53.
24. F. Rossi, M. Natasi, M. Cohen, C. Olsen, J. R. Tesmer and C. Egen, J. Mater. Res. 6 (1991) 1175.

**Figure Captions**

**Fig. 1** The GIXRD pattern of Si/C multilayers :- (a) as-prepared, (b) annealed at 500 °C, (c) Ar ion irradiated at fluence of $5 \times 10^{16}$ ions/cm$^2$, (d) same as in (c) and annealed at 500°C, (e) Ar ion irradiated at fluence of $1 \times 10^{17}$ ions/cm$^2$ and (f) same as in (e) and annealed at 500°C.

**Fig. 2** The Raman Spectra of Si/C multilayers in the range of 500 to 540 cm$^{-1}$:- (a) as-prepared, (b) annealed at 500 °C, (c) Ar ion irradiated at fluence of $5 \times 10^{16}$ ions/cm$^2$, (d) same as in (c) and annealed at 500°C, (e) Ar ion irradiated at fluence of $1 \times 10^{17}$ ions/cm$^2$ and (f) same as in (e) and annealed at 500°C.

**Fig. 3** The Raman Spectra of Si/C multilayers in the range 700 to 1100 cm$^{-1}$:- (a) as-prepared, (b) annealed at 500 °C, (c) Ar ion irradiated at fluence of $5 \times 10^{16}$ ions/cm$^2$, (d) same as in (c) and annealed at 500°C, (e) Ar ion irradiated at fluence of $1 \times 10^{17}$ ions/cm$^2$ and (f) same as in (e) and annealed at 500°C.

**Table 1** : Observed peak positions in GIXRD spectra of Si/C multilayer structures..

| Sample | 2θ Peak position | Phase |
|---|---|---|
| a) As deposited | 28.0, 51.30, 73.19 | Si |
| | 43.04, 64.02 | Hexagonal SiC |
| b) annealed (a) | 28.18, 51.10, 56.12, 73.15 | Si |
| | 42.96, 63.94 | Hexagonal SiC |
| c) low fluence irradiated | 27.97, 51.30, 73.27 | Si |
| | 43.12, 64.06 | Hexagonal SiC |
| e) annealed (c) | 27.70, 51.49, 73.30 | Si |
| | 43.22, 64.18 | Hexagonal SiC |
| f) high fluence irradiated | 37.13 | Hexagonal SiC |
| | 41.85, 59.59, 70.79 | Cubic SiC |
| g) annealed (f) | 51.11 | Si |
| | 60.30, 72.46 | Cubic SiC |
| | 38.43 | Hexagonal SiC |

**Table 1**

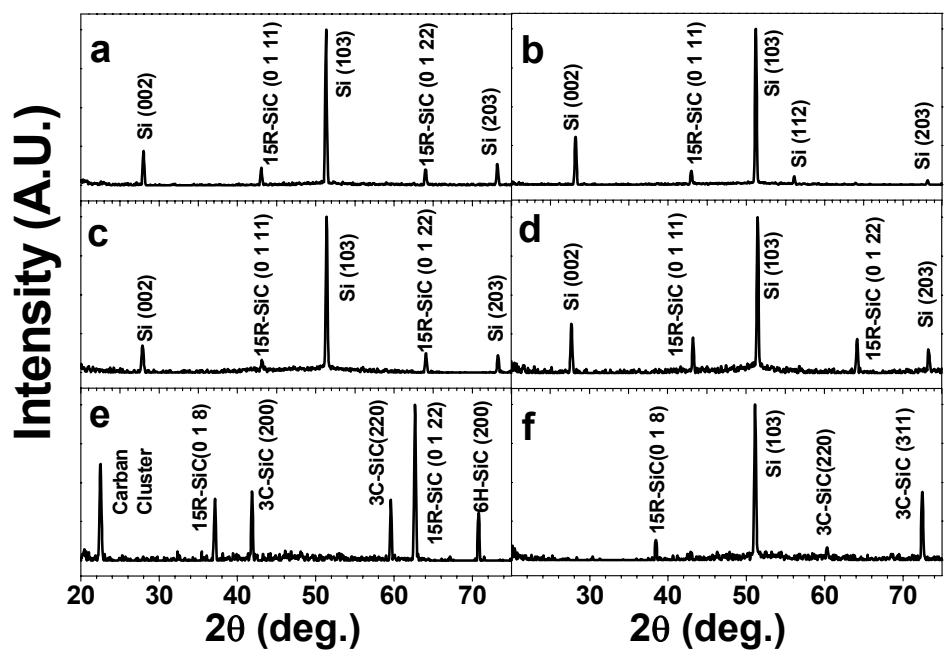

Fig. 1

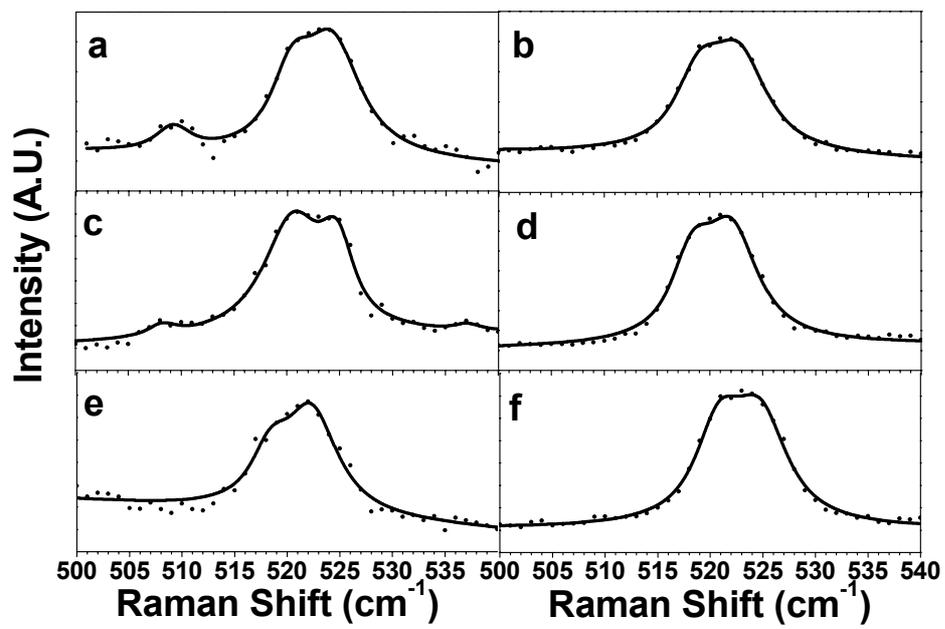

Fig. 2

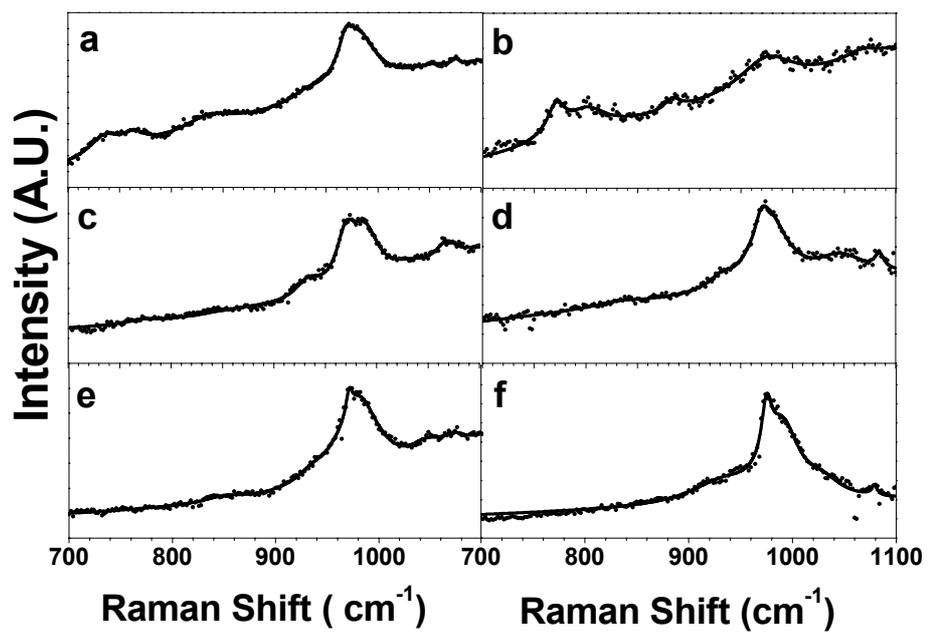

Fig. 3